\pdfoutput=1
\documentclass[twoside]{article}

\usepackage{graphicx}
\usepackage[scaled=.95]{helvet}
\usepackage[T1]{fontenc} 
\usepackage[left=1.5cm, right=1.5cm, top=1.785cm, bottom=2.0cm, columnsep=20pt]{geometry}
\usepackage{multicol} 
\usepackage[labelfont=bf,up,textfont=it,up]{caption} 
\usepackage{booktabs} 
\usepackage{float} 
\usepackage{hyperref} 

\usepackage{abstract} 

\usepackage{titlesec} 
\renewcommand\thesection{\Roman{section}} 
\titleformat{\section}[block]{\large\bfseries}{\thesection.}{1em}{} 

\title{\vspace{15mm}\fontsize{20pt}{10pt}\selectfont\textbf{Scattering Suppression and Absorption Enhancement in Contour Nanoantennas}\vspace{10mm}} 
\author{\Large{E. Doruk Onal}\thanks{Corresponding author: eonal@ku.edu.tr}{ , Kaan Guven}\\[5mm] 
\large Koc University \\ 
\vspace{2mm}}

\begin{document}
\maketitle 
\begin{abstract}
\vspace{5mm}
{\centering
\includegraphics[scale=1.3]{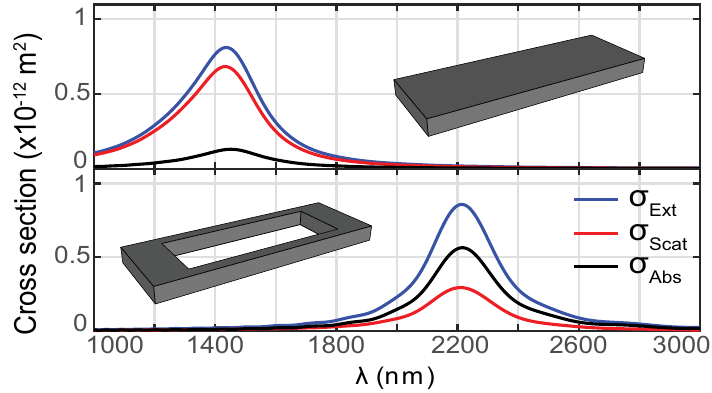}
\par
}
\vspace{5mm}
The expanding application spectrum of plasmonic nanoantennas demand versatile design approaches to tailor the antenna properties for specific requirements. The design efforts primarily concentrate on shifting the operation wavelength or enhancing the local fields by manipulating the size and shape of the nanoantenna.  Here, we propose a design path to control the absorption and scattering characteristics of a dipole nanoantenna by introducing a hollow region inside the nanostructure. The resulting contour geometry can significantly suppress the scattering of the dipole nanoantenna and enhance its absorption simultaneously. Both the dipole and the contour dipole nanoantenna couple to equivalent amount of the incident radiation. The dipole nanoantenna scatters 84\% of the coupled power (absorbs the remaining 16\%) whereas the contour dipole structure scatters only 28\% of the coupled power (absorbs the remaining 72\%). This constitutes the transformation from scatter to absorber nanoantenna. The scattering of a contour nanoantenna can be further suppressed by incorporating a lossless dielectric in the hollow region without altering its absorption. We also demonstrate the applicability of scattering suppression and absorption enhancement features of the contour design in other nanoantenna geometries such as the self-assembly compatible nanoantenna structures of nanodisk and nanoring chains. The benefits of the contour design can be readily utilized in diverse applications; including bioplasmonics, medical diagnosis/therapy, cloaked sensing, photovoltaics and thermoplasmonics.
\vspace{10mm}
\end{abstract}


\begin{multicols}{2} 
\section{Introduction}
Advances in nanotechnology paved the way to the miniaturization of antennas which provided the means to manipulate light at the subwavelength scale. The nanoantennas are being used in a wide range of applications: from sensing and cloaking to information processing, and even medical diagnosis and therapy.\cite{Gallinet2013,Muhlig2013,Butet2014} The working principle of the plasmonic nanoantennas is in their ability to confine light beyond the diffraction limit in the form of localized surface plasmon resonances (LSPRs).\cite{Agio2012} These resonances enhance the local electromagnetic field significantly and become very sensitive to changes in the local dielectric environment. 

Tailoring the properties of a nanoantenna for the vast variety of applications relies heavily on a careful design approach which involves three major parameters: shape, size and material. Nanoantennas in various shapes have been developed for different application purposes in the last two decades. Bowtie nanoantennas are designed to induce high local electromagnetic fields in between the sharp tips to be used in sensing applications; helical nanoantennas are employed in polarization conversion; multilayer dolmen nanostructures are shown to work as plasmonic rulers, and optical Yagi-Uda antennas are utilized to enhance and direct the emissions of a nanoemitter.\cite{Maksymov2012,Hofmann2007,Kosako2009,Dregely2012} The size of the nanoantenna primarily affects the operation wavelength and bandwidth. In radio frequency and microwave regimes, the dimensions of the antenna is directly related to its operation wavelength. However, in the optical regime, the field penetration into the metal body is significant due to the excitation of localized surface plasmons. This field penetration leads to an effective resonance wavelength which takes into account the optical properties of the metal $L=\lambda_{eff}/2$.\cite{Biagioni2011,Novotny2007} The material of choice for optical nanoantennas has been dominated by gold and silver due to their small Ohmic losses and inertness under atmospheric conditions. However, there is a growing interest to incorporate new materials for better fabrication characteristics and enhanced heat stability.\cite{Boltasseva2014,Guler2014,Naik2013,Adams2011} The losses in metallic nanoantennas are significant and inevitable. Considerable amount of research has focused on circumventing this problem by introducing a gain medium, employing superconductors, or constructing all-dielectric nanoantennas.\cite{Soukoulis2014, Khurgin2015} On the other hand, the lossy nature of the metals can be put into use in devices specifically designed for absorption purposes such as solar energy harvesting\cite{Wu2012}, metamaterial perfect absorbers \cite{Rhee2014} and high optical field generators for near-field microscopy.\cite{Nien2013} 

Nanoantennas that absorb a significant proportion of the incident radiation are utilized in the recently emerging field of thermoplasmonics which exploits the Joule heating caused by the plasmon resonances.\cite{Baffou2013a,Baffou2013,Hao2011} The resultant nanoscale heat sources found use in biological applications such as selective photothermal cancer cell treatment \cite{Stern2008,Hirsch2003,Tsai2013} and light controlled plasmonic drug delivery.\cite{Xiong2014} In chemistry, absorber nanoantennas act as nanosources of heat, electrons and strong optical near-fields to control reactions in a nanoscale precision.\cite{Baffou2014} Designing highly absorptive plasmonic structures for thermoplasmonic applications initiated interest in refractory plasmonic materials to account for the melting point depreciation in nanoscale structures.\cite{Jiang2003,Li2014} Absorber nanoantennas are also employed in photovoltaic systems to trap solar radiation in the form of localized surface plasmons, which generate strong near-fields and enhance the creation of electron-hole pairs in the semiconductor.\cite{Atwater2010} Similarly, in solar thermophotovoltaic devices, instead of creating electron-hole pairs, absorber nanoantennas convert sunlight into heat which via a thermal emitter is coupled to the photovoltaic cell for power generation.\cite{Bermel2011} Theoretically, the heat generated in the plasmonic nanostructures under continuous illumination is directly related to the absorption cross section $(\sigma_{Abs}$) and the incident light intensity (I): $P = \sigma_{Abs} \times I$.\cite{Baffou2009,Baffou2014a}

Plasmonic nanoantennas are also widely utilized in sensing applications. Free-standing nanoantennas resonant in the biological transparency window ($650-1350 \; nm$) serve as real-time, in-situ spectroscopy for in depth investigation of cellular processes.\cite{Adato2013} There is also a growing interest in the self-assembly of nanoantennas, especially inside living cells, by using nanoparticles attached to polymer or biomolecule templates, in particular when the resulting nanoantenna system is too large to go through the cell membrane. \cite{Tan2011} In-vitro diagnostic devices based on arrays of plasmonic nanoantennas have also been reported.\cite{Cetin2014} In recent years, considerable effort is focused on cloaking the nanoantenna sensors to decrease the scattering and disturbance on the probed environment.\cite{Alu2010,Monticone2014,Fleury2014,Novotny2011} Unlike ideal cloaking where the presence of an object is completely shielded from the surrounding electromagnetic field, in sensor cloaking the aim is to minimize the scattering caused by the nanoantenna while preserving the coupling to the incident radiation.\cite{Fleury2014a} This concept of ``seeing without being seen'' \cite{Alu2009} has potential applications in communication technologies and in biological/medical imaging by lowering noise and interference.\cite{Chen2012} 

In view of the aforementioned applications, designing low-scattering nanoantennas that has strong coupling to the incident field is crucial for increasing the efficiency in sensing applications. On the other hand, thermoplasmonic and photovoltaic applications can benefit considerably from enhancing the absorption cross section of a nanoantenna. In this paper, we demonstrate the scattering suppression and absorption enhancement in plasmonic nanoantennas by introducing a hollow region into the structure, which results in a contour geometry. While several contour nanoantenna structures have been studied in the past for shifting resonances to achieve deep subwavelength light modulation, for manipulating the damping of the plasmonic resonances, and for improving refractive index sensing capabilities \cite{Lecarme2014,Chau2013,Sederberg2011,Gharibi2014}, our study exposes the significant freedom gained by a simple contour design to extend and tune the scattering and absorption properties of a nanoantenna, without altering its outer dimensions. To the best of our knowledge, these features have not been exploited systematically before.
\vspace{5mm}
\begin{figure}[H]
\centering
  \includegraphics[scale=1]{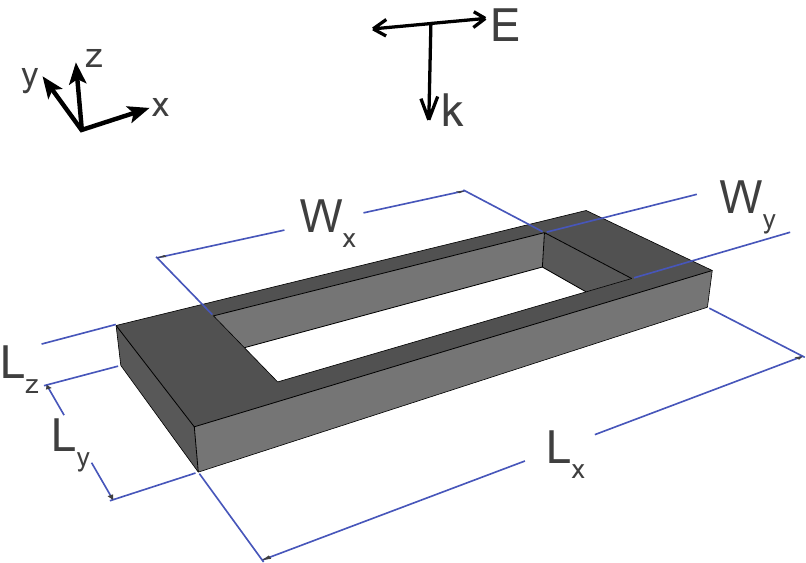}
  \caption{A schematic view of the contour dipole antenna under linearly polarized plane wave illumination with outer nanoantenna ($L_{x} \times L_{y} \times L_{z}$) and hollow ($W_{x} \times W_{y}$) size parameters.}
  \label{fgr:fig1}
\end{figure}

The rest of the article is structured in 5 sections: The first section provides a comparative study on the scattering suppression and absorption enhancement in dipole and contour dipole nanoantennas. In the second section, we present the effects of the contour geometry on the transmission/reflection properties of a periodic array of contour nanoantennas. The third section gives a qualitative description of the mechanisms providing the scattering suppression and the absorption enhancement based on the local electric field and the induced current distributions of the contour nanoantenna. In the fourth section, the effects of a dielectric-filled hollow on the scattering and absorption of the nanoantenna is discussed. The last section highlights the general applicability of the contour design by illustrating its use in self-assembling nanoantenna structures.

\section{Single Nanoantenna Extinction, Scattering \& Absorption Cross-Sections}

The structures used in this study are referred to as dipole nanoantennas since only the fundamental electric dipole mode is excited under linearly polarized plane-wave illumination. We present a numerical study of silver nanoantennas in near-infrared region. The simulations are performed using the Lumerical FDTD software in the wavelength range of 1000 - 3000 nm. Free standing nanoantennas without any substrate are studied to eliminate complications such as resonance shift and broadening. In all of the performed simulations the nanoantenna height ($L_{z}$) is fixed at 30 nm. Nanoantennas of varying length ($L_{x} = 300 - 700 \; nm$) and width ($L_{y} = 20 - 250 \; nm$) are studied. All contour dipole nanoantenna designs are based on the same parent dipole structure ($L_{x} \times L_{y} \times L_{z} = 400 \times 100 \times 30 \; nm$) with varying hollow dimensions ($W_{x} = 100 - 360 \; nm, \; W_{y} = 20 - 80 \; nm$). For modelling the complex refractive index of silver, we use Brendal-Bormann model which is shown to be in good agreement with experimental observations in the studied wavelength range.\cite{Rakic1998,Jahanshahi2014,Brendel1992}
\vspace{5mm}
\begin{figure}[H]
\centering
  \includegraphics[scale=1]{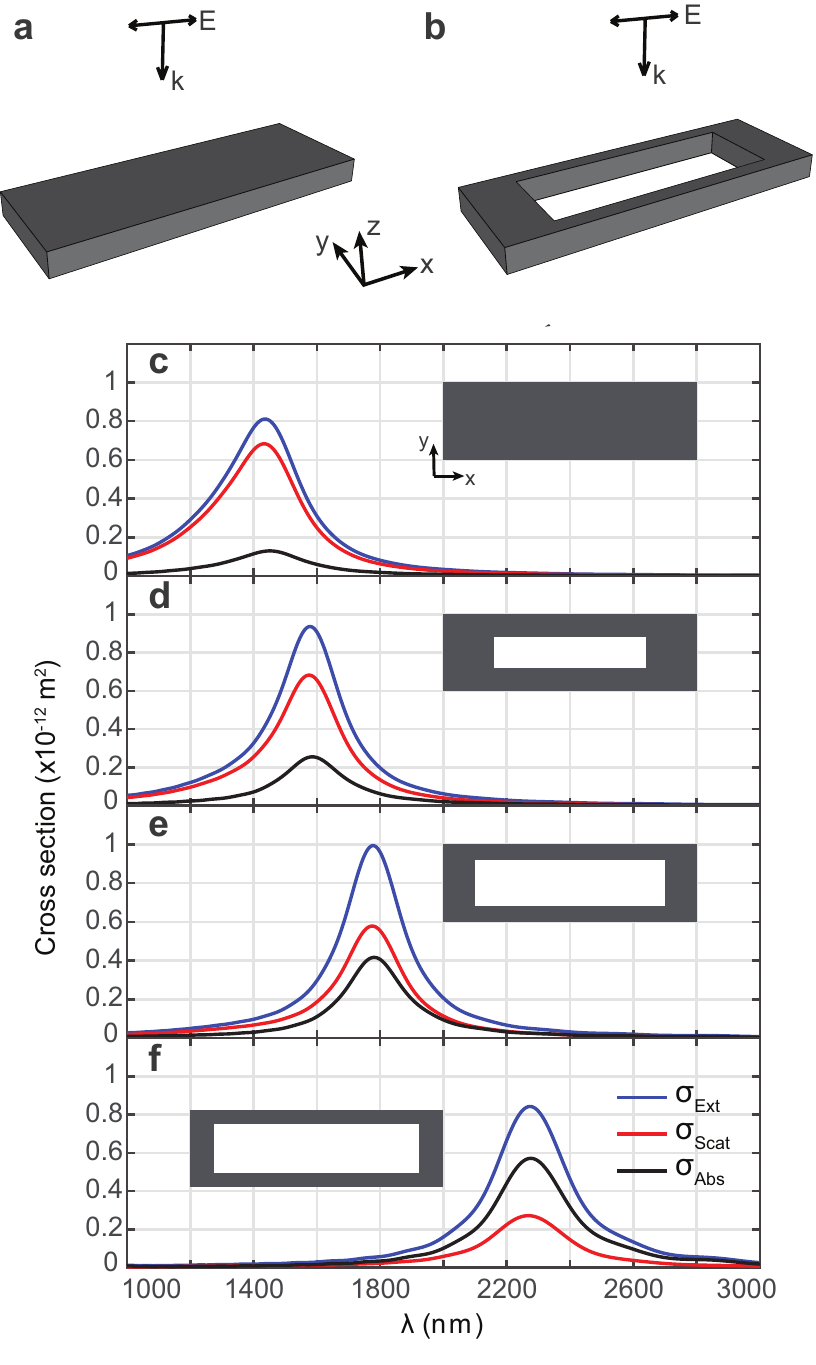}
  \caption{(a-b) A schematic view of the plain dipole and the contour dipole antenna. The scattering, absorption and extinction cross section of the dipole (c) and the contour dipole nanoantennas (d-f). The outer dimensions of the nanoantennas are the same: $L_{x} \times L_{y} \times L_{z} = 400 \times 100 \times 30 \: nm$. The contour size $W_{x} \times W_{y}$ changes  by (d) $240 \times 40 \: nm$, (e) $300 \times 60 \: nm$ (f) $360 \times 80 \: nm$, as illustrated by the insets.}
  \label{fgr:fig2}
\end{figure}

The extinction cross section ($\sigma_{Ext}$) of an object is a measure of its effective shadowing area and calculated by: $\sigma_{Ext} = P_{tot}/I$ where $P_{tot}$ is the total energy loss from the incident beam and $I$ is the incident light intensity. One of the contributions to this loss of intensity is due to redirection of the propagating beam which is related to the scattering cross section: $\sigma_{Scat} = P_{scat}/I$ where $P_{scat}$ represents scattered power by the object. Due to energy conservation the rest of the power loss must be due to the absorption of the object and by definition the absorption cross section is: $\sigma_{Abs} = \sigma_{Ext} - \sigma_{Scat}$.\cite{Dahlin2012} We calculated the scattering and extinction cross sections ($\sigma_{Scat} \; \& \; \sigma_{Ext}$) of the dipole and contour dipole nanoantennas using the FDTD simulations and consequently obtained the absorption cross section ($\sigma_{Abs}$).

Comparing the extinction, scattering and absorption cross sections of a dipole and three contour dipole nanoantennas (of the same outer dimensions) revealed the changes induced by the presence and varying size of the hollow region (Fig 2d-f): The resonance wavelength is red-shifted with increasing hollow size. As the outer dimensions remain intact, this resonance shift enables deeper subwavelength control of light by contour dipole nanoantenna ($L_{contour}=400 \; nm=\lambda_{res}/5.5$) compared to a dipole nanoantenna of the same size ($L_{dipole}=400 \; nm=\lambda_{res}/3.5$). Increasing the hollow size transforms the nanoantenna from scatterer to absorber in character, as the absorption surpasses the scattering cross section (Fig 2e, f). Furthermore, the extinction cross section remains roughly constant throughout the crossover of the scattering and absorption cross-sections. This implies that, by starting with a proper sized dipole nanoantenna, one can carve out a hollow region to produce a contour nanoantenna with tailored scattering and absorption cross sections, at the desired operation wavelength, while preserving the extinction cross section.

Conventional design approaches for manipulating the resonance wavelength and the scattering behavior of dipole nanoantennas focus on dimensional alterations of the structure. To contrast the effects of the contour geometry, we present a comparative study on the effects of changing the length and the width of a dipole nanoantenna on the resonance wavelength, relative scattering efficiency and scattering, absorption and extinction cross sections of a dipole nanoantenna (Fig 3). Figure 3a, c \& e show the effects of changing the nanoantenna length ($L_{x}$), at a fixed width ($L_{y} = 100 \: nm$). The resonance wavelength as well as the scattering and absorption cross sections change linearly whereas the relative scattering efficiency ($RSE = \sigma_{Scat} / \sigma_{Ext}$) remains constant with the changing nanoantenna length. We introduced RSE as a measure of the extent of scattering in the extinction cross section. An RSE smaller than 0.5 indicates to an absorber nanoantenna, contrarily a scatterer nanoantenna has an RSE larger than 0.5.

Figure 3b, d \& f illustrate the effects of changing the nanoantenna width ($L_{y}$), at a fixed length ($L_{x} = 400 \: nm$). Significant variation of the relevant parameters is achieved only for narrow nanoantennas ($L_{y} < 50 \; nm$). We will compare the narrow dipole nanoantennas to the contour dipole structures later in this section.

\begin{figure}[H]
\centering
  \includegraphics[scale=1]{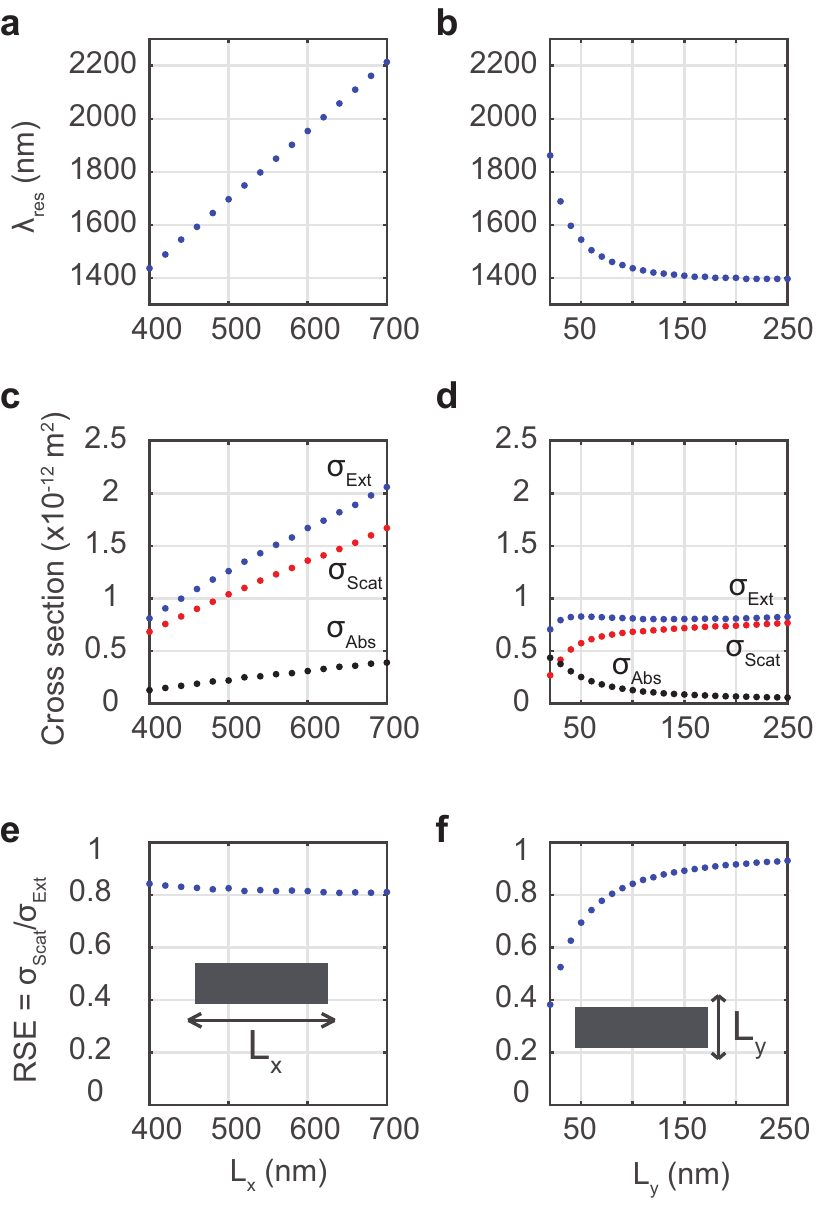}
  \caption{Effect of changing dipole nanoantenna length-$L_{x}$ (first column) and width-$L_{y}$ (second column) on the resonance wavelength (a, b), $\sigma_{Scat}, \; \sigma_{Abs} \; \& \; \sigma_{Ext}$ (c, d), and the relative scattering efficiency ($RSE = \sigma_{Scat}/\sigma_{Ext}$). Insets illustrate alterations in nanoantenna length and width.}
  \label{fig3}
\end{figure}

The control over the RSE of a dipole nanoantenna based on length and width alterations is limited to narrow structures only (Fig 4a). Figure 4 illustrates the extent of freedom gained by the contour design approach in modifying the RSE of a parent dipole nanoantenna (marked by the black rectangle with dimensions: $ L_{x} \times L_{y} = 400 \times 100 \: nm $). The dipole nanoantenna acts as a scatterer ($RSE > 0.5$) almost in the entire range of the axes. This observation is also consistent with the calculations based on Mie theory for nanospheres and nanoshells which indicates that small particles ($ L \leq \sim\lambda/20 $) are predominantly absorbers whereas larger particles act as scatterers.\cite{Grady2004}

The RSE of the contour dipole nanoantenna can be suppressed from 0.84 for the parent dipole (marked by the black rectangle in Fig 4a) to 0.32 by enlarging the hollow region (Fig 4b). Note that when $W_{y} > 70 \; nm$, the RSE can be brought below 0.5 and the selected constant RSE curves are running horizontally. Evidently, the width ($W_{y}$) rather than the length ($W_{x}$) of the hollow region governs the change in the RSE. We will elucidate this dependence later on while discussing the local field and current density of the contour dipole antenna.

\begin{figure}[H]
\centering
  \includegraphics[scale=1]{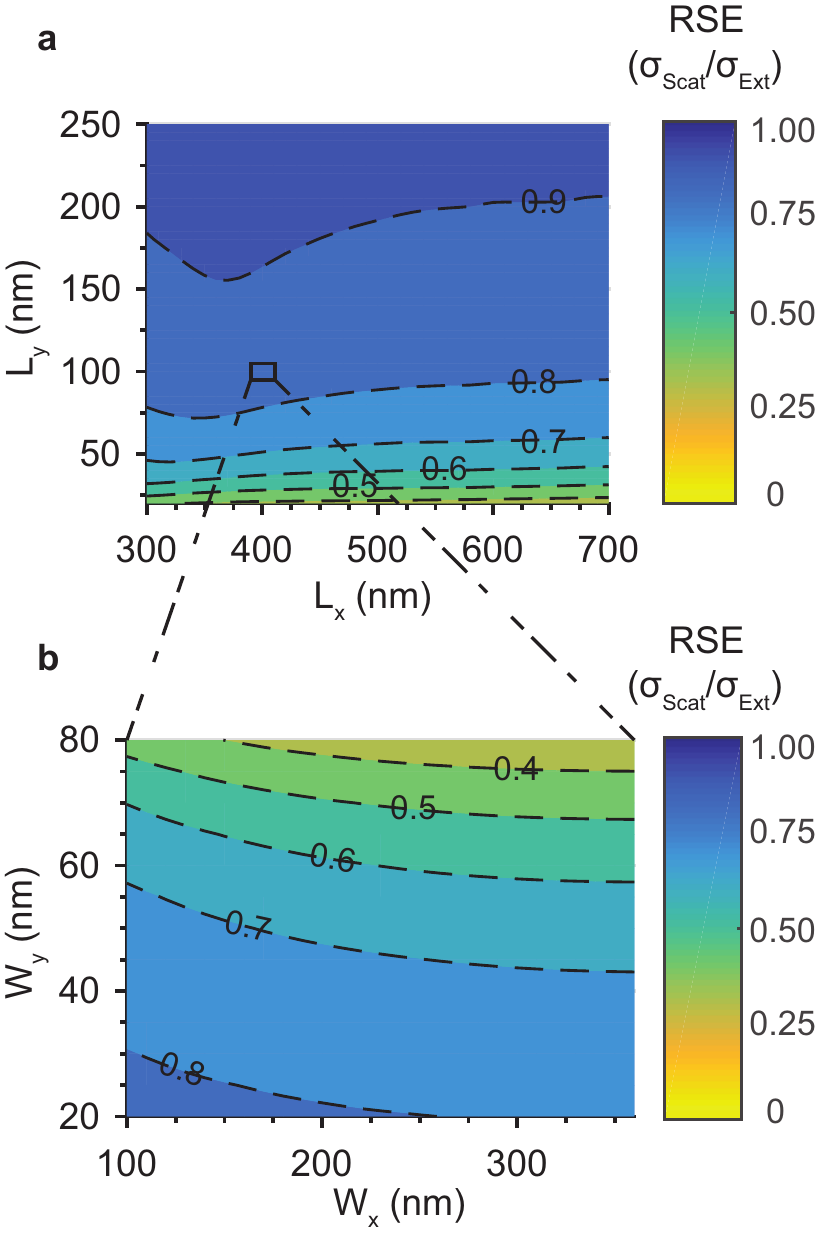}
  \caption{(a) RSE of a dipole nanoantenna as a function of its length ($L_{x}$) and width ($L_{y}$). (b) The expanded RSE of a contour dipole nanoantenna which is based on the parent dipole with dimensions: $L_{x} \times L_{y} = 400 \times 100 \: nm$ (marked by the black rectangular region in (a)). RSE > (<) 0.5 indicates predominantly scatterer (absorber) nanoantennas.}
  \label{fig4}
\end{figure}

We conclude this section by summarizing the modification of the scattering, absorption and extinction cross sections of the dipole and the contour dipole nanoantenna by their respective design parameters as shown in the first and second columns of Figure 5. The black rectangular regions in the figures of the first column, mark the location of the parent dipole structure ($ L_{x} \times L_{y} = 400 \times 100 \: nm$) of the contour dipole nanoantennas in the second column. Thus, the second column of figures can be seen as the expansion of the rectangular region through the contour design and signifies the additional range gained in the corresponding parameters.

The similarity of the scattering and extinction cross section profiles of the dipole nanoantenna (Fig 5a, e) emphasizes its scattering character. The vertical running contour curves in Fig 5a, e indicate that the properties of the dipole nanoantenna are mainly governed by its length and the nanoantenna width becomes effective only below 50 nm. The absorption cross section remains small and relatively insensitive to the changes in the design parameters (Fig 5c).

The significance of the contour nanoantenna design is in the scattering and absorption cross section profiles which exhibit inverse and comparable variations as a function of contour parameters (Fig 5b, d). Consequently, the extinction cross section remains almost constant (Fig 5f). The horizontal running contour curves indicate that the cross sections are dictated by the width of the hollow. To demonstrate the scatterer to absorber transition in contour dipole nanoantennas, we plot the cross sections as a function of hollow width ($W_{y}$) at fixed hollow length ($W_{x} = 300 \: nm$). Figure 5g clearly shows the crossing of $\sigma_{Scat} \; \& \; \sigma_{Abs}$.

In cloaked sensing applications, suppression of scattering without compromising on the coupling to the incident field is essential. The contour nanoantenna design constitute a viable path for scattering suppression. The dashed arrows in Fig 5a, b indicate the directions to be taken to suppress the scattering in the dipole and contour dipole nanoantennas. The arrows in the lower panels, mark the trend in absorption and extinction cross sections (Fig 5c-f) along the scattering suppression path for each nanoantenna respectively. The scattering suppression path for the dipole nanoantenna fails to preserve the extinction cross section. However, the contour dipole nanoantenna preserves the extinction cross section which represents the coupling to the incident field. The utilization of contour dipole nanoantennas in cloaked sensing applications will be discussed in more detail in the later sections.
\vspace{11cm}
\begin{figure*}[p]
\centering
  \includegraphics[scale=1]{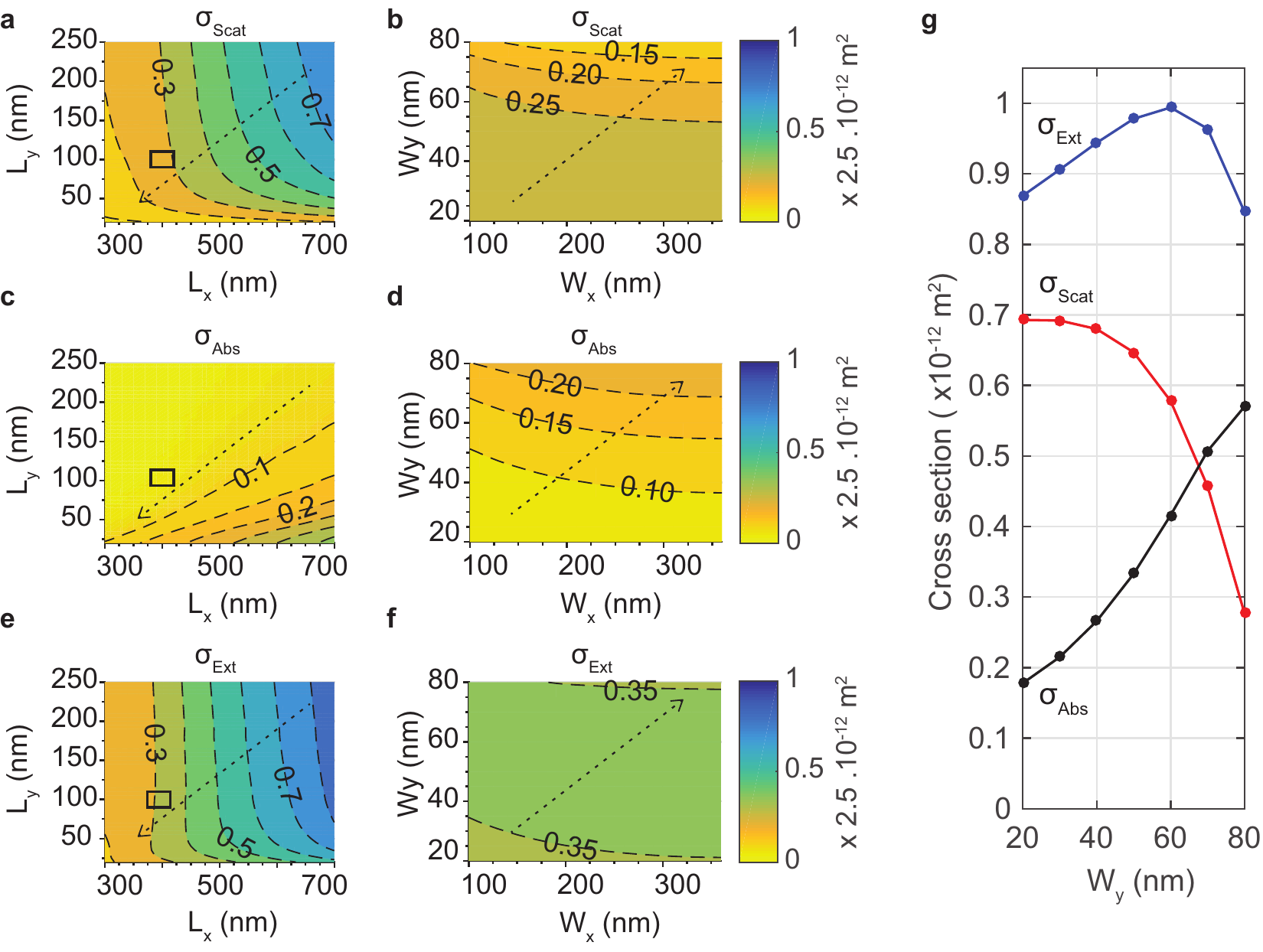}
  \caption{The scattering (a), absorption (c) and extinction (e) cross sections of the dipole nanoantenna as a function of its length ($L_{x}$) and width ($L_{y}$). The black rectangles indicate the location of the parent dipole ($ L_{x} \times L_{y} = 400 \times 100 \: nm$) which is used to construct the contour dipole nanoantenna. The scattering (b), absorption (d) and (f) extinction cross sections of the contour dipole antenna as a function of hollow length ($W_{x}$) and width ($W_{y}$). The dashed arrow in panels (a-f) indicate the direction for scattering suppression. (g) The scattering , absorption and extinction cross sections of the contour dipole nanoantenna as a function of contour width at fixed contour length $W_{x} = 300 \: nm$.}
  \label{fig5}
\end{figure*}

\section{Periodic Nanoantenna Array Transmission, Reflection \& Absorption}
\vspace{-3mm}
\begin{figure}[H]
\centering
  \includegraphics[scale=1]{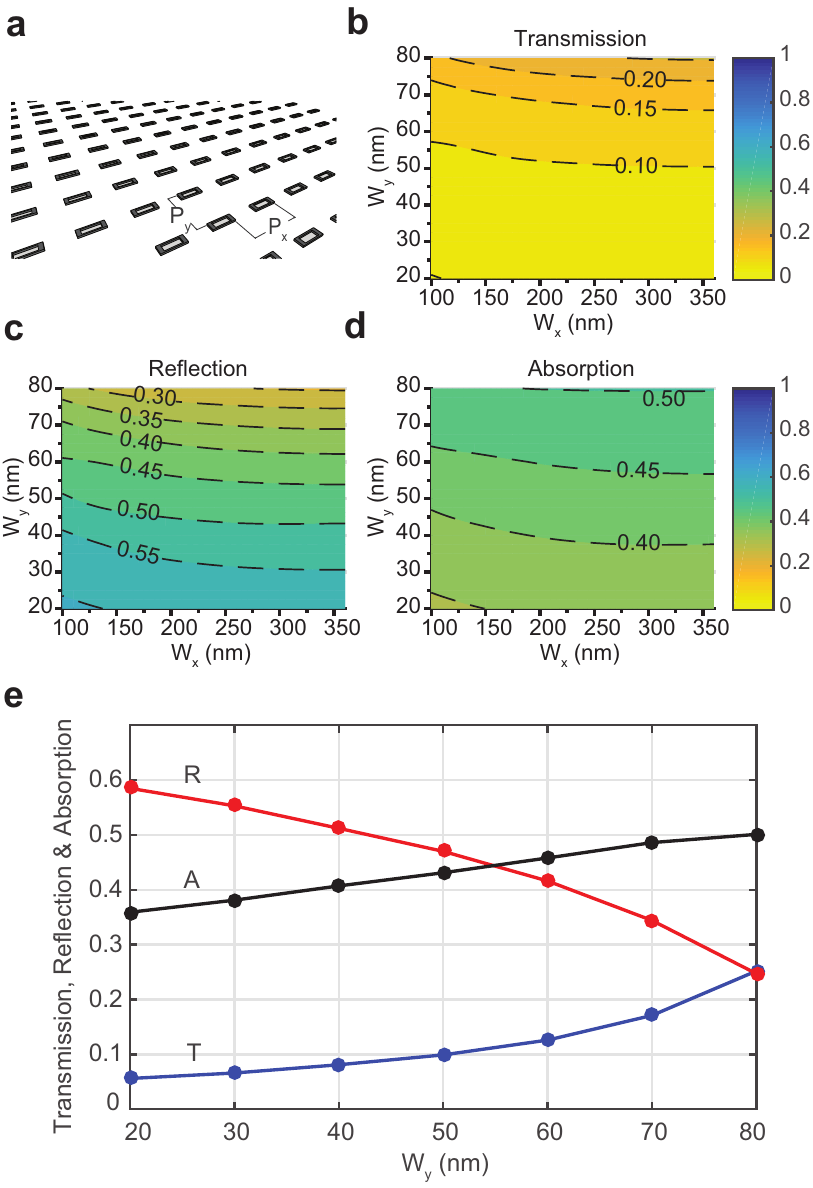}
  \caption{(a) The schematic of the periodic contour dipole nanoantenna array. The transmission (b), reflection (c) and absorption (d) of the nanoantenna array as a function of the hollow length ($W_{x}$) and width ($W_{y}$). (e) Demonstration of the reflection/absorption crossover by widening the hollow at a fixed length $W_{x} =300 \: nm$. Periodicity in both directions ($P_{x} = P_{y} = 800 \: nm$) and parent dipole nanoantenna dimensions ($L_{x} \times L_{y} \times L_{z} = 400 \times 100 \times 30 \: nm$) are fixed.}
  \label{fig6}
\end{figure}

Periodic nanoantenna arrays provide a functional layer with application-specific reflection, transmission and absorption properties and their usage is already reported in optical nanocircuits\cite{Dregely2011}, sensors \cite{Ng2011} and photovoltaics.\cite{Simovski2013} The electromagnetic response of the nanoantenna array is primarily determined by that of the unit cell (constituent nanoantenna). The refractive index of the substrate and the periodicity of the array account for the secondary factors that influence the electromagnetic response of the nanoantenna array.\cite{Feuillet-Palma2013,Yan2014} In order to relate to the results presented in the previous section and to focus on the effects of the contour design, we consider a free-standing contour nanoantenna array with a typical periodicity of $P_{x} = P_{y} = 800 \: nm$ (Fig 6a). The transmission (T) and reflection (R) under normal incidence are calculated from the FDTD simulation and the absorption (A) is determined by the relation: $A = 1 - (T + R)$. The extinction (E) is given by $E = R + A = 1 - T$. 

Figures 6 b, c \& d show the transmission, reflection and absorption as a function of the hollow dimensions. In accordance with the results of a single contour dipole antenna (Fig 5g), the periodic array exhibits a crossover from reflective to absorptive behavior (Fig 6e).  Unlike the response of the single contour dipole nanoantenna, the extinction (transmission) of the periodic array shows a decreasing (increasing) trend with increasing hollow width ($W_{y}$).

\section{Scattering Suppression \& Absorption Enhancement Mechanisms}

For a qualitative understanding of the mechanism behind the scattering suppression and the absorption enhancement, we investigate the electric field and the current density distribution of the dipole and the contour dipole nanoantennas at their respective resonance wavelengths. The dominant electric field component along the direction of incident polarization ($E_{x}$) shows the electric dipole moments induced in each nanostructure (Fig 7a, b). 

The electric fields at both ends of the dipole nanoantenna are oriented in the same (-x) direction, which defines a dipole moment indicated by the blue arrow (Fig 7a). The contour nanoantenna possesses an additional opposing dipole moment (red arrow) which is induced across the hollow region (Fig 7b). Since the scattering (i.e the radiated energy) of the nanoantenna is proportional to the square of the net dipole moment, reducing the net dipole moment suppresses the scattering of the contour nanoantenna. A similar scattering suppression is utilized in plasmonic cloaking of sensors, where the dipole moment of a dielectric nanoantenna is reduced by the opposite polarization induced in the surrounding plasmonic cloak with negative permittivity ($P_{cloak} = (\epsilon_{cloak} - \epsilon_{0}) E_{incident} $). 

The enhancement of absorption can be inferred from the on-resonance current density profiles of the nanostructures. The contour geometry restricts the surface current flow to narrow channels whereas the current flow is spread over the entire surface of the dipole nanoantenna (Fig 7e, f). As the contour geometry has considerably smaller cross sectional area ($1/5$ of the dipole), by Pouillet's law ($R = \rho L/A$) its electrical resistance increases by the same amount.\cite{Griffiths2013} Applying the Ampere's law on a rectangular path at the yz-plane outlining each nanoantenna at their respective resonances, reveals that the dipole and the contour dipole nanoantenna have a similar amount of current flow (Fig 7g). The same amount of current passing through a region of increased resistance increases the Ohmic losses ($P = I^2 R$), and significantly enhances the absorption in the contour dipole structure. Note that, the absorption cross section remains relatively unaffected by the hollow length change, whereas the increasing hollow width increases the absorption significantly (Fig 5e).

\begin{figure}[H]
\centering
  \includegraphics[scale=1]{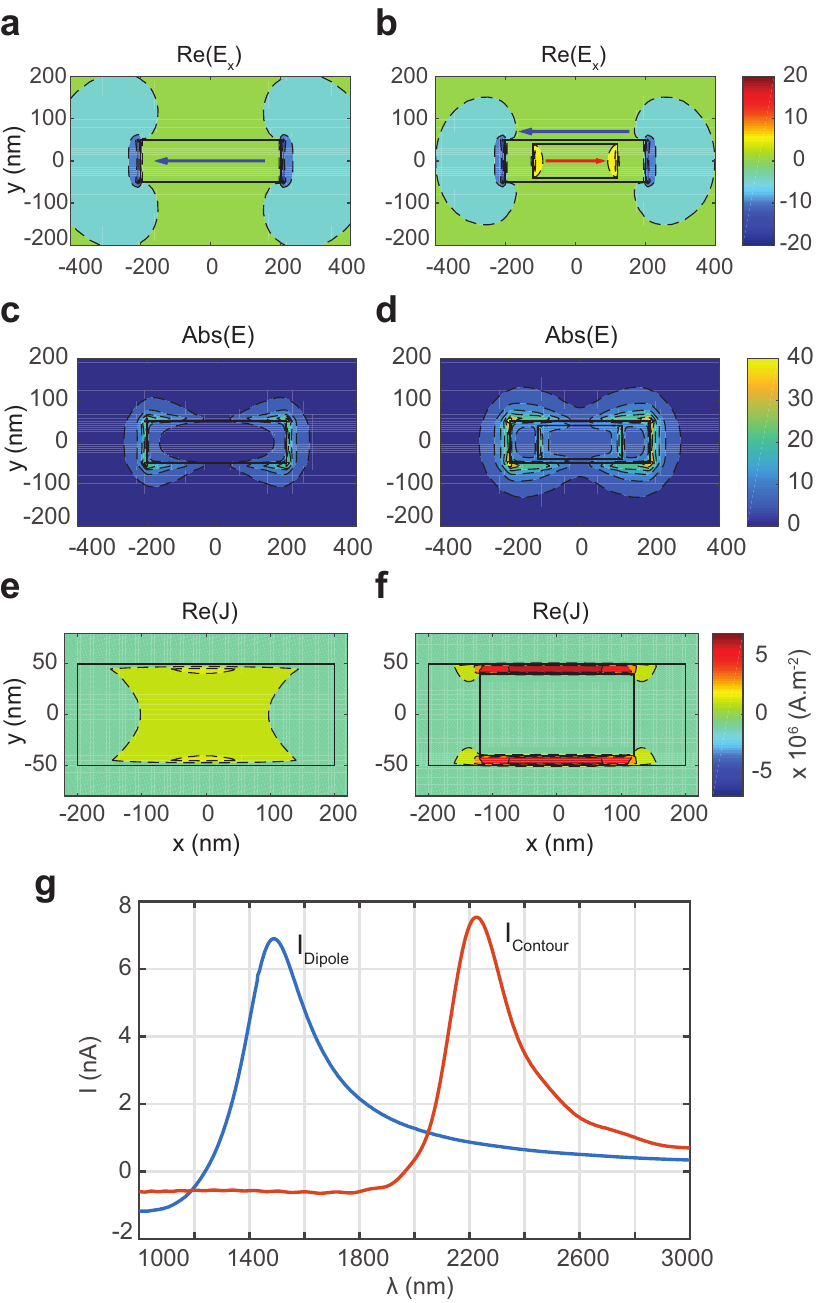}
  \caption{
	Comparison of the (a, b) real part of the x-component of the electric field, (c, d) electric field magnitude, (e, f) the current density for the dipole (first column) and the contour dipole (second column)  nanoantenna. The nanoantennas have the same outer dimensions ($L_{x} \times L_{y} \times L_{z} = 400 \times 100 \times 30 \: nm$) and the contour dipole has a hollow region ($W_{x} \times W_{y} = 240 \times 80 \: nm$). The arrows in (a, b) indicate the induced electric dipole moments.(g) The total current integrated through the cross section of the dipole nanoantenna (blue curve) and the contour dipole nanoantenna (red curve). All figures are plotted at the respective resonance of the nanoantennas.}
  \label{fig7}
\end{figure}

The binary effect of the hollow as a scattering suppressor and absorption enhancer in the contour dipole nanoantennas, makes these structures promising candidates for noninvasive sensing and communication applications. Scattering suppression reduces the noise and the disturbance in the environment of measurement whereas increased absorption preserves the extinction cross section and maintains strong coupling to the incident radiation. The induced strong local fields increase the interaction with matter for near-field spectroscopic applications (Fig 7c, d). 

\section{Further Scattering Suppression by Changing the Refractive Index of the Hollow}

Increasing the refractive index of the hollow region by incorporating a dielectric medium enables the construction of contour dipole nanoantennas with even lower relative scattering efficiencies (Fig 8).

\begin{figure}[H]
\centering
  \includegraphics[scale=1]{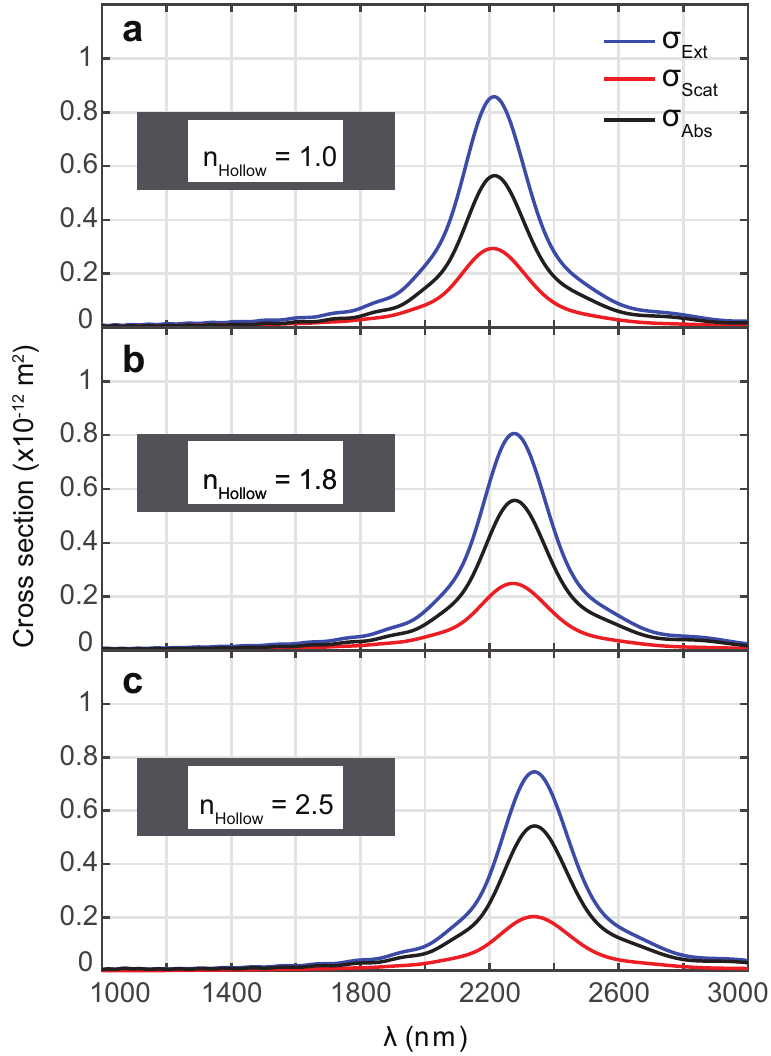}
  \caption{The effect of a lossless dielectric medium with refractive index ($n_{Hollow}$), filling the hollow region of the contour dipole on the scattering, absorption and extinction cross section spectra: (a) $n_{Hollow}$=1 (reference), (b) $n_{Hollow}$ = 1.8, (c) $n_{Hollow}$ = 2.5, as illustrated by the insets. The contour dipole has the same design as given in Fig.7.}
  \label{fig8}
\end{figure}

Unlike the changes corresponding to the dimensional alterations of the hollow, the scattering suppression induced by increasing the refractive index of the dielectric ($n_{Hollow}$) preserves the absorption rather than the extinction cross section (Fig 9a). Conservation of absorption cross section while suppressing the scattering of a nanoantenna adds another functionality to the contour design and enables the independent manipulation of scattering from absorption cross section. The RSE of the contour dipole nanoantenna can be further reduced (Fig 9b) and the resonance wavelength shifts towards slightly larger wavelengths (Fig 9c).

The scattering suppression mechanism in contour dipole nanoantennas is attributed to the cancellation of antiparallel polarization vectors induced across the nanoantenna ($P_{Ant}=(\epsilon_{Ant} - \epsilon_0)E_{incident} $) and the hollow ($ P_{Hollow}=(\epsilon_{Hollow} - \epsilon_0)E_{incident} $) that result in a diminished net polarization ($P_{net} = P_{Ant} - P_{Hollow} $).\cite{Alu2005} Increasing the refractive index of the hollow ($ n_{Hollow} = \sqrt{\epsilon_{Hollow}} $) produces a larger polarization vector inside the hollow dielectric and further reduces the net polarization of the contour dipole structure. The reduced net dipole moment causes the additional depreciation in scattering cross section. On the other hand, absorption cross section remains unaffected as the cross sectional area for the flow of resonating charges is preserved. Changing the hollow refractive index constitutes an alternative design path to further suppress the scattering of a dipole nanoantenna without enhancing its absorption.

\begin{figure}[H]
\centering
  \includegraphics[scale=1]{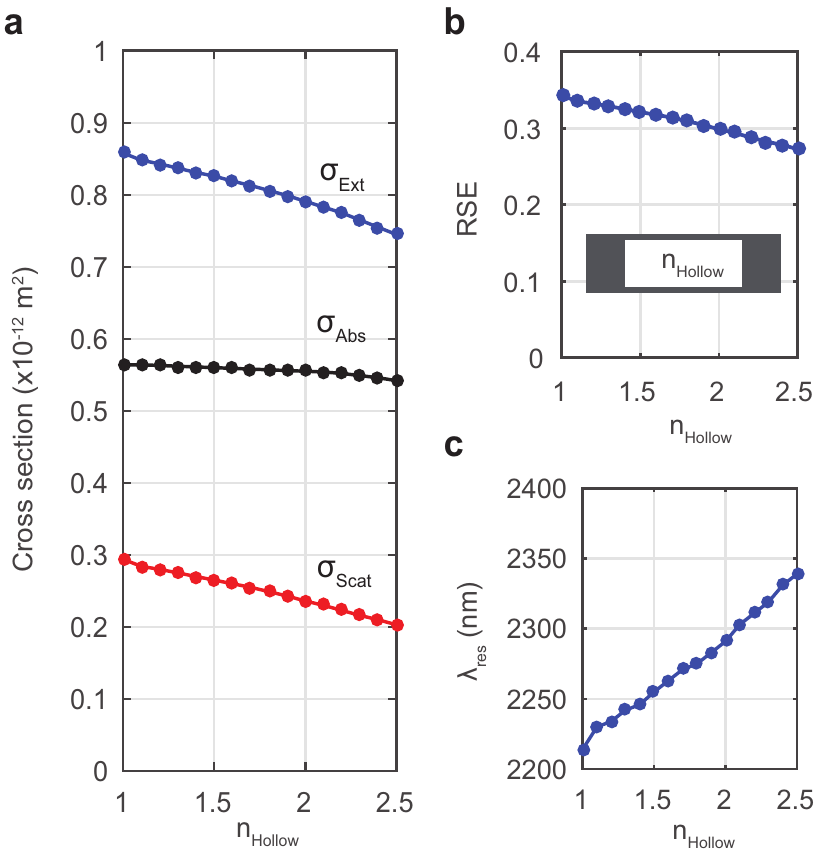}
  \caption{(a) The effect of the refractive index of the medium filling the hollow region on (a) the extinction, absorption and scattering cross sections, (b) the relative scattering efficiency and (c) the resonance wavelength of the contour dipole antenna. Inset illustrates the contour nanoantenna geometry and the hollow refractive index ($n_{Hollow}$). The antenna dimensions are as given in Fig.7.}
  \label{fig9}
\end{figure}

\section{Scattering Suppression in Self-Assembling Nanoantennas}

With the advancing self-assembly techniques, more and more research is focused on producing self-assembling counterparts of top-down constructed plasmonic nanoantennas.\cite{Gao2012,Klinkova2014,Slaughter2012} Replacing a dipole nanoantenna with a chain of nanoparticles achieves light modulation in deeper subwavelength regime by shifting the resonance to longer wavelengthes while preserving the size of the structure.\cite{Atay2004} This resonance shift is attributed to the prolonged path for charge oscillations on the surface of the nanoparticle chain. Replacing the solid nanoparticles with nanoshells increases the resonance wavelength further. The oscillating electrons are restricted to move along the even longer path on the outer surface of the nanoshells.\cite{Li2014a}

\begin{figure}[H]
\centering
  \includegraphics[scale=1]{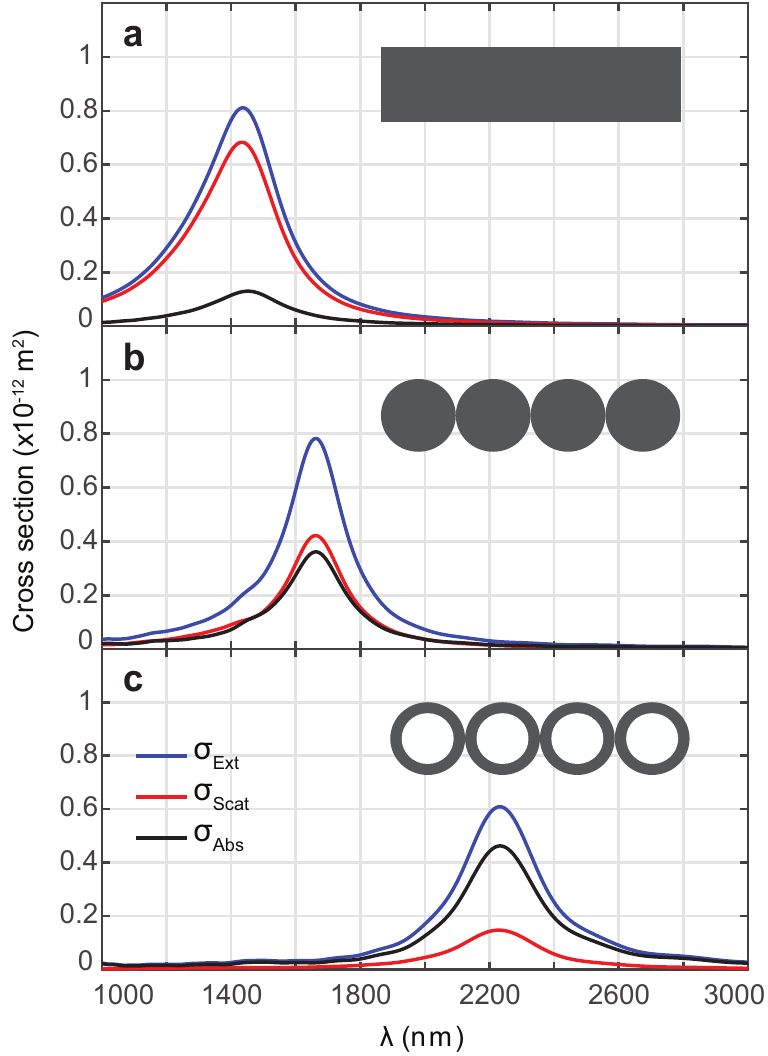}
  \caption{The extinction, scattering and absorption spectra of (a) the dipole antenna ($L_{x} \times L_{y} \times L_{z} = 400 \times 100 \times 30 \; nm$) (b) an assembly of four identical nanodisks ($R_{disk} \times h_{disk} = 100 \times 30 \; nm$) (c) an assembly of four identical nanorings ($R_{out} \times R_{in} \times h_{disk} = 100 \times 70 \times 30 \; nm$). The nanodisk chain (middle inset) can be considered as a dipole nanoantenna (top inset) contoured from the outside and the nanoring chain (bottom inset) can be considered as a dipole nanoantenna contoured from both the outside and the inside.}
  \label{fig10}
\end{figure}

In addition to changing the resonance wavelength, replacing dipole nanoantennas with nanodisk and nanoring chains causes a significant scattering suppression and absorption enhancement (Fig 10a-c). Relative scattering efficiency is dramatically reduced from 0.84 for the dipole nanoantenna to 0.57 for the nanodisk chain and further to 0.28 for the nanoring chain. Apart from being promising candidates for subwavelength light modulation, nanoring chains significantly suppress scattering and enable the construction of self-assembling, deep subwavelength, absorber nanoantennas. These self-assembling, low-scattering nanoantenna systems also has great potential for cloaked sensing applications especially in biological systems to monitor cellular processes where current cloaked sensor designs are either too large to pass through the cell membrane or too difficult to construct inside the cell using the available self-assembly techniques.

\section{Conclusions}
We propose the contour antenna geometry as a design scheme to tailor the scattering and absorption properties of the dipole nanoantennas. The primary feature of this design is to suppress the scattering of the nanoantenna which can be utilized in low-scattering sensors to reduce noise in optical communication and spectroscopic applications. Absorption enhancement is another feature of the contour geometry that is beneficial in photovoltaic and thermoplasmonic applications where the absorption of the electromagnetic field is critical.

The absorption enhancement and the scattering suppression occur simultaneously in contour nanoantennas causing little variation in the extinction cross section. In other words, the contour geometry determines how the extinction is distributed between the scattering and absorption cross sections. A predominantly scatterer nanoantenna can be transformed to become absorber in character, while the antenna extinction is preserved ($RSE_{Dipole} = 0.84, \; RSE_{Contour} = 0.28$). This provides the optimal conditions for cloaked sensing applications where reduced scattering yet strong coupling to the external field is required.

Deeper subwavelength light manipulation is another feature of the contour dipole nanoantenna as it resonates at a longer wavelength than its parent dipole. Therefore, the resonance-matching dipole nanoantenna needs to be significantly larger than the contour structure. Incorporating different dielectric materials in the hollow suppresses the scattering further without altering the absorption cross section which comprises a different functionality than modifying the size of the hollow region.

The contour design is applicable to other nanoantenna geometries together with the conclusions drawn for the dipole nanostructure. The scattering of a nanodisk chain, which is a self-assembled counterpart of the dipole nanoantenna, can be suppressed substantially by replacing the nanodisks with nanorings ($RSE_{Nanodisk} = 0.57, \; RSE_{Nanoring} = 0.28$). With the readily available features, the contour design can be implemented in various nanoantenna geometries to be employed in a wide range of applications.



\end{multicols}


\begin{thebibliography}{10}

\bibitem{Gallinet2013}
B.~Gallinet and O.~J.~F. Martin, ``{Refractive index sensing with subradiant
  modes: A framework to reduce losses in plasmonic nanostructures},'' {\em ACS
  Nano}, vol.~7, no.~8, pp.~6978--6987, 2013.

\bibitem{Muhlig2013}
S.~M\"{u}hlig, A.~Cunningham, J.~Dintinger, M.~Farhat, S.~B. Hasan, T.~Scharf,
  T.~B\"{u}rgi, F.~Lederer, and C.~Rockstuhl, ``{A self-assembled
  three-dimensional cloak in the visible.},'' {\em Sci. Rep.}, vol.~3, p.~2328,
  2013.

\bibitem{Butet2014}
J.~Butet and O.~J.~F. Martin, ``{Refractive index sensing with Fano resonant
  plasmonic nanostructures: a symmetry based nonlinear approach},'' {\em
  Nanoscale}, vol.~6, no.~24, pp.~15262--15270, 2014.

\bibitem{Agio2012}
M.~Agio, ``{Optical antennas as nanoscale resonators},'' {\em Nanoscale},
  vol.~4, no.~3, p.~692, 2012.

\bibitem{Maksymov2012}
I.~S. Maksymov, I.~Staude, A.~E. Miroshnichenko, and Y.~S. Kivshar, ``{Optical
  Yagi-Uda nanoantennas},'' {\em Nanophotonics}, vol.~1, no.~1, pp.~65--81,
  2012.

\bibitem{Hofmann2007}
H.~F. Hofmann, T.~Kosako, and Y.~Kadoya, ``{Design parameters for a
  nano-optical Yagi-Uda antenna},'' {\em New J. Phys.}, vol.~9, no.~07,
  pp.~1--12, 2007.

\bibitem{Kosako2009}
T.~Kosako, H.~F. Hofmann, and Y.~Kadoya, ``{Directional emission of light from
  a nano-optical Yagi-Uda antenna},'' {\em Nat. Photonics}, vol.~4, no.~March,
  p.~4, 2009.

\bibitem{Dregely2012}
D.~Dregely, K.~Lindfors, J.~Dorfm\"{u}ller, M.~Hentschel, M.~Becker,
  J.~Wrachtrup, M.~Lippitz, R.~Vogelgesang, and H.~Giessen, ``Plasmonic
  antennas, positioning, and coupling of individual quantum systems,'' {\em
  Phys. Status Solidi B}, vol.~249, no.~4, pp.~666--677, 2012.

\bibitem{Biagioni2011}
P.~Biagioni, J.-S. Huang, and B.~Hecht, ``Nanoantennas for visible and infrared
  radiation,'' {\em Rep. Prog. Phys.}, vol.~75, no.~2, p.~024402, 2012.

\bibitem{Novotny2007}
L.~Novotny, ``{Effective wavelength scaling for optical antennas},'' {\em Phys.
  Rev. Lett.}, vol.~98, no.~26, pp.~1--4, 2007.

\bibitem{Boltasseva2014}
A.~Boltasseva, ``{Empowering plasmonics and metamaterials technology with new
  material platforms},'' {\em MRS Bull.}, vol.~39, no.~05, pp.~461--468, 2014.

\bibitem{Guler2014}
U.~Guler, V.~M. Shalaev, and A.~Boltasseva, ``{Nanoparticle plasmonics: going
  practical with transition metal nitrides},'' {\em Mater. Today (Oxford, U.
  K.)}, vol.~00, no.~00, pp.~1--11, 2014.

\bibitem{Naik2013}
G.~V. Naik, V.~M. Shalaev, and A.~Boltasseva, ``{Alternative plasmonic
  materials: Beyond gold and silver},'' {\em Adv. Mater.}, vol.~25, no.~24,
  pp.~3264--3294, 2013.

\bibitem{Adams2011}
D.~C. Adams, S.~Inampudi, T.~Ribaudo, D.~Slocum, S.~Vangala, N.~a. Kuhta, W.~D.
  Goodhue, V.~a. Podolskiy, and D.~Wasserman, ``{Funneling light through a
  subwavelength aperture with epsilon-near-zero materials},'' {\em Phys. Rev.
  Lett.}, vol.~107, no.~13, pp.~1090--1099, 2011.

\bibitem{Soukoulis2014}
C.~M. Soukoulis, T.~Koschny, P.~Tassin, N.-H. Shen, and B.~Dastmalchi, ``{What
  is a good conductor for metamaterials or plasmonics},'' {\em Nanophotonics},
  vol.~0, no.~0, pp.~1--6, 2014.

\bibitem{Khurgin2015}
J.~B. Khurgin, ``{How to deal with the loss in plasmonics and metamaterials},''
  {\em Nat. Nanotechnol.}, vol.~10, no.~1, pp.~2--6, 2015.

\bibitem{Wu2012}
C.~Wu, B.~{Neuner III}, J.~John, A.~Milder, B.~Zollars, S.~Savoy, and
  G.~Shvets, ``{Metamaterial-based integrated plasmonic absorber/emitter for
  solar thermo-photovoltaic systems},'' {\em J. Opt.}, vol.~14, no.~2,
  p.~024005, 2012.

\bibitem{Rhee2014}
J.~Rhee, Y.~Yoo, K.~Kim, Y.~Kim, and Y.~Lee, ``Metamaterial-based perfect
  absorbers,'' {\em Journal of Electromagnetic Waves and Applications},
  vol.~28, no.~13, pp.~1541--1580, 2014.

\bibitem{Nien2013}
L.~W. Nien, S.~C. Lin, B.~K. Chao, M.~J. Chen, J.~H. Li, and C.~H. Hsueh,
  ``{Giant electric field enhancement and localized surface plasmon resonance
  by optimizing contour bowtie nanoantennas},'' {\em J. Phys. Chem. C},
  vol.~117, no.~47, pp.~25004--25011, 2013.

\bibitem{Baffou2013a}
G.~Baffou, P.~Berto, E.~Berm\'{u}dez Ure\~{n}a, R.~Quidant, S.~Monneret,
  J.~Polleux, and H.~Rigneault, ``Photoinduced heating of nanoparticle
  arrays,'' {\em ACS Nano}, vol.~7, no.~8, pp.~6478--6488, 2013.

\bibitem{Baffou2013}
G.~Baffou and R.~Quidant, ``{Thermo-plasmonics: Using metallic nanostructures
  as nano-sources of heat},'' {\em Laser Photonics Rev.}, vol.~7, no.~2,
  pp.~171--187, 2013.

\bibitem{Hao2011}
J.~Hao, L.~Zhou, and M.~Qiu, ``Nearly total absorption of light and heat
  generation by plasmonic metamaterials,'' {\em Phys. Rev. B}, vol.~83,
  p.~165107, Apr 2011.

\bibitem{Stern2008}
J.~M. Stern, J.~Stanfield, W.~Kabbani, J.~T. Hsieh, and J.~a. Cadeddu,
  ``{Selective prostate cancer thermal ablation with laser activated gold
  nanoshells},'' {\em Journal of Urology}, vol.~179, no.~2, pp.~748--753, 2008.

\bibitem{Hirsch2003}
L.~R. Hirsch, R.~J. Stafford, J.~A. Bankson, S.~R. Sershen, B.~Rivera, R.~E.
  Price, J.~D. Hazle, N.~J. Halas, and J.~L. West, ``Nanoshell-mediated
  near-infrared thermal therapy of tumors under magnetic resonance guidance,''
  {\em Proc. Natl. Acad. Sci. U. S. A.}, vol.~100, no.~23, pp.~13549--13554,
  2003.

\bibitem{Tsai2013}
M.~F. Tsai, S.~H.~G. Chang, F.~Y. Cheng, V.~Shanmugam, Y.~S. Cheng, C.~H. Su,
  and C.~S. Yeh, ``{Au nanorod design as light-absorber in the first and second
  biological near-infrared windows for in vivo photothermal therapy},'' {\em
  ACS Nano}, vol.~7, no.~6, pp.~5330--5342, 2013.

\bibitem{Xiong2014}
W.~Xiong, R.~Mazid, L.~W. Yap, X.~Li, and W.~Cheng, ``{Plasmonic caged gold
  nanorods for near-infrared light controlled drug delivery},'' {\em
  Nanoscale}, vol.~6, no.~23, pp.~14388--14393, 2014.

\bibitem{Baffou2014}
G.~Baffou, J.~Polleux, H.~Rigneault, and S.~Monneret, ``{Super-heating and
  micro-bubble generation around plasmonic nanoparticles under cw
  illumination},'' {\em J. Phys. Chem. C}, vol.~118, no.~9, pp.~4890--4898,
  2014.

\bibitem{Jiang2003}
Q.~Jiang, S.~Zhang, and M.~Zhao, ``{Size-dependent melting point of noble
  metals},'' {\em Mater. Chem. Phys.}, vol.~82, no.~1, pp.~225--227, 2003.

\bibitem{Li2014}
W.~Li, U.~Guler, N.~Kinsey, G.~V. Naik, A.~Boltasseva, J.~Guan, V.~M. Shalaev,
  and A.~V. Kildishev, ``{Refractory Plasmonics with Titanium Nitride:
  Broadband Metamaterial Absorber},'' {\em Adv. Mater.}, vol.~26, no.~47,
  pp.~7959--7965, 2014.

\bibitem{Atwater2010}
H.~a. Atwater and A.~Polman, ``{Plasmonics for improved photovoltaic
  devices.},'' {\em Nat. Mater.}, vol.~9, no.~3, pp.~205--213, 2010.

\bibitem{Bermel2011}
P.~Bermel, M.~Ghebrebrhan, M.~Harradon, Y.~Yeng, I.~Celanovic, J.~D.
  Joannopoulos, and M.~Soljacic, ``{Tailoring photonic metamaterial resonances
  for thermal radiation},'' {\em Nanoscale Res. Lett.}, vol.~6, no.~1, p.~549,
  2011.

\bibitem{Baffou2009}
G.~Baffou, R.~Quidant, and C.~Girard, ``{Heat generation in plasmonic
  nanostructures: Influence of morphology},'' {\em Appl. Phys. Lett.}, vol.~94,
  no.~15, pp.~1--3, 2009.

\bibitem{Baffou2014a}
G.~Baffou and R.~Quidant, ``{Nanoplasmonics for chemistry.},'' {\em Chem. Soc.
  Rev.}, vol.~43, no.~11, pp.~3898--907, 2014.

\bibitem{Adato2013}
R.~Adato and H.~Altug, ``{In-situ ultra-sensitive infrared absorption
  spectroscopy of biomolecule interactions in real time with plasmonic
  nanoantennas.},'' {\em Nat. Commun.}, vol.~4, p.~2154, 2013.

\bibitem{Tan2011}
S.~J. Tan, M.~J. Campolongo, D.~Luo, and W.~Cheng, ``{Building plasmonic
  nanostructures with DNA.},'' {\em Nat. Nanotechnol.}, vol.~6, no.~5,
  pp.~268--276, 2011.

\bibitem{Cetin2014}
A.~E. Cetin, A.~F. Coskun, B.~C. Galarreta, M.~Huang, D.~Herman, A.~Ozcan, and
  H.~Altug, ``{Handheld high-throughput plasmonic biosensor using computational
  on-chip imaging},'' {\em Light: Sci. Appl.}, vol.~3, no.~1, p.~e122, 2014.

\bibitem{Alu2010}
A.~Al\`u and N.~Engheta, ``Cloaking a receiving antenna or a sensor with
  plasmonic metamaterials,'' {\em Metamaterials}, vol.~4, no.~2–3, pp.~153 --
  159, 2010.

\bibitem{Monticone2014}
F.~Monticone and A.~Al\`u, ``Do cloaked objects really scatter less?,'' {\em
  Phys. Rev. X}, vol.~3, p.~041005, Oct 2013.

\bibitem{Fleury2014}
R.~Fleury, J.~Soric, and A.~Al\`u, ``Physical bounds on absorption and
  scattering for cloaked sensors,'' {\em Phys. Rev. B}, vol.~89, p.~045122, Jan
  2014.

\bibitem{Novotny2011}
L.~Novotny, ``{From near-field optics to optical antennas feature},'' {\em
  Phys. Today}, no.~July, pp.~47--52, 2011.

\bibitem{Fleury2014a}
R.~Fleury and A.~Al{\`u}, ``Cloaking and invisibility: A review,'' in {\em
  Forum for Electromagnetic Research Methods and Application Technologies
  (FERMAT)}, vol.~1, pp.~1--24, 2014.

\bibitem{Alu2009}
A.~Al\`u and N.~Engheta, ``Cloaking a sensor,'' {\em Phys. Rev. Lett.},
  vol.~102, p.~233901, Jun 2009.

\bibitem{Chen2012}
P.-Y. Chen, J.~Soric, and A.~Al\`u, ``Invisibility and cloaking based on
  scattering cancellation,'' {\em Adv. Mater.}, vol.~24, no.~44,
  pp.~OP281--OP304, 2012.

\bibitem{Lecarme2014}
O.~Lecarme, Q.~Sun, K.~Ueno, and H.~Misawa, ``{Robust and Versatile Light
  Absorption at Near-Infrared Wavelengths by Plasmonic Aluminum Nanorods},''
  {\em ACS Photonics}, vol.~1, no.~6, pp.~538--546, 2014.

\bibitem{Chau2013}
Y.~F. Chau, W.~H. Lin, M.~J. Sung, C.~Y. Jheng, S.~C. Jheng, and D.~P. Tsai,
  ``{Numerical Investagation of a Castle-like Contour Plasmonic Nanoantenna
  with Operating Wavelengths Ranging in Ultraviolet-Visible, Visible Light, and
  Infrared Light},'' {\em Plasmonics}, vol.~8, no.~2, pp.~755--761, 2013.

\bibitem{Sederberg2011}
S.~Sederberg and a.~Y. Elezzabi, ``{Nanoscale plasmonic contour bowtie antenna
  operating in the mid-infrared},'' {\em Opt. Express}, vol.~19, no.~16,
  p.~15532, 2011.

\bibitem{Gharibi2014}
M.~Gharibi, H.~Khoshsima, B.~Olyaeefar, and S.~Khorram, ``{Field enhancement by
  plasmonic contour H-shaped nano-antenna},'' {\em Eur. Phys. J. D}, vol.~68,
  no.~5, pp.~2--6, 2014.

\bibitem{Rakic1998}
A.~D. Rakic, A.~B. Djuri\v{s}ic, J.~M. Elazar, and M.~L. Majewski, ``{Optical
  Properties of Metallic Films for Vertical-Cavity Optoelectronic Devices},''
  {\em Appl. Opt.}, vol.~37, no.~22, p.~5271, 1998.

\bibitem{Jahanshahi2014}
P.~Jahanshahi, M.~Ghomeishi, and F.~R.~M. Adikan, ``{Study on dielectric
  function models for surface Plasmon resonance structure},'' {\em Sci. World
  J.}, vol.~2014, p.~503749, 2014.

\bibitem{Brendel1992}
R.~Brendel and D.~Bormann, ``{An infrared dielectric function model for
  amorphous solids},'' {\em Journal of Applied Physics}, vol.~71, no.~1,
  pp.~1--6, 1992.

\bibitem{Dahlin2012}
A.~Dahlin, {\em Plasmonic biosensors an integrated view of refractometric
  detection}.
\newblock Amsterdam Washington DC: Ios Press, 2012.

\bibitem{Grady2004}
N.~K. Grady, N.~J. Halas, and P.~Nordlander, ``{Influence of dielectric
  function properties on the optical response of plasmon resonant metallic
  nanoparticles},'' {\em Chem. Phys. Lett.}, vol.~399, no.~1-3, pp.~167--171,
  2004.

\bibitem{Dregely2011}
D.~Dregely, R.~Taubert, J.~Dorfm\"{u}ller, R.~Vogelgesang, K.~Kern, and
  H.~Giessen, ``{3D optical Yagi-Uda nanoantenna array.},'' {\em Nat. Commun.},
  vol.~2, p.~267, 2011.

\bibitem{Ng2011}
B.~Ng, S.~M. Hanham, V.~Giannini, Z.~C. Chen, M.~Tang, Y.~F. Liew, N.~Klein,
  M.~H. Hong, and S.~a. Maier, ``{Lattice resonances in antenna arrays for
  liquid sensing in the terahertz regime},'' {\em Opt. Express}, vol.~19,
  no.~15, p.~14653, 2011.

\bibitem{Simovski2013}
C.~Simovski, D.~Morits, P.~Voroshilov, M.~Guzhva, P.~Belov, and Y.~Kivshar,
  ``Enhanced efficiency of light-trapping nanoantenna arrays for thin-film
  solar cells,'' {\em Opt. Express}, vol.~21, pp.~A714--A725, Jul 2013.

\bibitem{Feuillet-Palma2013}
C.~Feuillet-Palma, Y.~Todorov, A.~Vasanelli, and C.~Sirtori, ``{Strong near
  field enhancement in THz nano-antenna arrays.},'' {\em Sci. Rep.}, vol.~3,
  p.~1361, 2013.

\bibitem{Yan2014}
C.~Yan and O.~J.~F. Martin, ``Periodicity-induced symmetry breaking in a fano
  lattice: Hybridization and tight-binding regimes,'' {\em ACS Nano}, vol.~8,
  no.~11, pp.~11860--11868, 2014.

\bibitem{Griffiths2013}
D.~Griffiths, {\em Introduction to electrodynamics}.
\newblock Boston: Pearson, 2013.

\bibitem{Alu2005}
A.~Al\`{u} and N.~Engheta, ``{Achieving transparency with plasmonic and
  metamaterial coatings},'' {\em Phys. Rev. E: Stat., Nonlinear, Soft Matter
  Phys.}, vol.~72, no.~1, pp.~1--9, 2005.

\bibitem{Gao2012}
B.~Gao, G.~Arya, and A.~R. Tao, ``{Self-orienting nanocubes for the assembly of
  plasmonic nanojunctions},'' {\em Nat. Nanotechnol.}, vol.~7, no.~7,
  pp.~433--437, 2012.

\bibitem{Klinkova2014}
A.~Klinkova, H.~Th\'{e}rien-Aubin, A.~Ahmed, D.~Nykypanchuk, R.~M. Choueiri,
  B.~Gagnon, A.~Muntyanu, O.~Gang, G.~C. Walker, and E.~Kumacheva, ``Structural
  and optical properties of self-assembled chains of plasmonic nanocubes,''
  {\em Nano Lett.}, vol.~14, no.~11, pp.~6314--6321, 2014.

\bibitem{Slaughter2012}
L.~S. Slaughter, B.~a. Willingham, W.~S. Chang, M.~H. Chester, N.~Ogden, and
  S.~Link, ``{Toward plasmonic polymers},'' {\em Nano Lett.}, vol.~12, no.~8,
  pp.~3967--3972, 2012.

\bibitem{Atay2004}
T.~Atay, J.-H. Song, and A.~V. Nurmikko, ``Strongly interacting plasmon
  nanoparticle pairs: From dipole-dipole interaction to conductively coupled
  regime,'' {\em Nano Lett.}, vol.~4, no.~9, pp.~1627--1631, 2004.

\bibitem{Li2014a}
Z.~Li, S.~Butun, and K.~Aydin, ``Touching gold nanoparticle chain based
  plasmonic antenna arrays and optical metamaterials,'' {\em ACS Photonics},
  vol.~1, no.~3, pp.~228--234, 2014.

\end{thebibliography}
\end{document}